\documentclass[twocolumn,prl]{revtex4}
\usepackage{float}
\usepackage{makeidx}
\usepackage{graphicx,amsmath}

\begin{document}

\title{Controllable coupled-resonator-induced transparency in a dual-recycled Michelson
interferometer}
\author{Xudong Yu, Wei Li, Yuanbin Jin,  Jing Zhang$^{\dagger}$}
\affiliation{The State Key Laboratory of Quantum Optics and
Quantum Optics Devices, Institute of
Opto-Electronics, Synergetic Innovation Center of Extreme Optics,\\
Shanxi University, Taiyuan 030006, P.R.China \label{in}}

\begin{abstract}

We theoretically and experimentally study the effect of the
coupled-resonator-induced transparency in the Michelson
interferometer with the dual-recycled configuration, which is
equivalent to the adjustable coupled resonator. The coupling
strength, corresponding to the splitting of the reflection spectrum and associated with the width of the
coupled-resonator-induced transparent window, can be controlled by
adjusting the arm lengths, i.e., the relative phase of the
interference arms on the 50/50 beam splitter. Thus, this tunability of
the coupling strength can be very fast, and the absorptive and dispersive
properties can be effectively controlled. This work provides a new
system to coherently control and storage optical field.

\end{abstract}

\maketitle
The destructive interference between the excitation pathways in the
three level atomic system, which is well known as the
electromagnetically induced transparency (EIT)
\cite{harris,Marangos,Fleischhauer}, has been widely studied and has
attracted many attentions in the past twenty years since the first
experiment was reported \cite{Boller}. The steep dispersion and the
low absorption at the EIT window make it become the primary choice
for coherent optical information storage and freeze the light
\cite{hau,Kash,Budker,Liu1,philips}, even storing quantum state in
quantum information process \cite{Honda,Appel,Yan,xu}.

The similar coherence and interference phenomenon, named as EIT-like
effect, has been demonstrated in the classical systems such as
plasma \cite{harris2,Litvak,Shvets}, mechanical or electric
oscillators \cite{Hemmer,Garrido}, coupled optical resonators
\cite{Opatrny-2001,Smith-2004,Smith-2004-1,Maleki-2004,Yanik-2004,Di},
optical parametric oscillators \cite{ma,ye1,ye2}, optomechanical
systems \cite{Weis,Safavi,marquardt} and even some metamaterial configurations \cite{fan,jenkins,zhu}. Especially, the analog of EIT in coupled
optical resonators, also called coupled-resonator-induced
transparency, has made great progress in experiment, which has been
reported in the different optical systems, such as compound glass
waveguide platform using relatively large resonators \cite{Chu}, fiber
ring resonators \cite{Dumeige},
coupled fused-silica microspheres \cite{Naweed-2005,Totsuka},
integrated micron-size silicon optical resonator systems
\cite{Xu-2006,Li,lanyang,lipson}, photonic crystal cavities \cite{Yang} and graphene-ring resonators \cite{shubin}. Two dynamically tuned resonators for
stopping light have been proposed theoretically
\cite{Maleki-2004,Yanik-2004,Otey} and demonstrated experimentally
\cite{Xu-2007}. These works provide the new ways of utilizing
coupled optical resonators for the optical communication and the
simulation of coherent effect in quantum optics. In this paper, we
present a new optical system to realize coupled-resonator-induced
transparency, in which the Michelson interferometer with the
dual-recycled configuration is employed. This system is equivalent
to the coupled resonators with tunable coupling strength. The absorptive and dispersive
properties can be effectively controlled by adjusting the arm lengths, i.e., the relative phase of the
interference arms on the 50/50 beam splitter. This method can realize the very fast tuning of the coupling strength in the coupled-resonator-induced transparency. It would be useful for the capture, storage, and release of the light, even quantum field.

\begin{figure}
\centerline{
\includegraphics[width=2.5in]{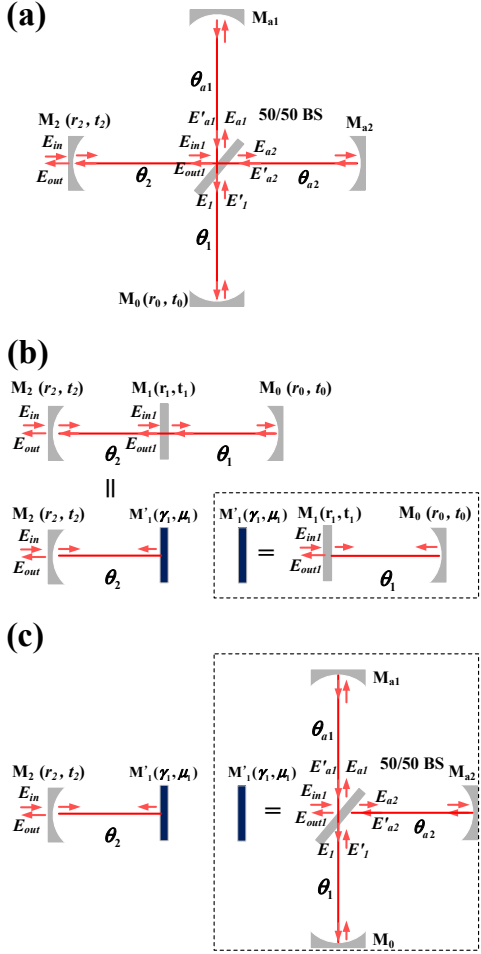}
} \vspace{0.1in}
\caption{(Color online). (a) The schematic diagram of dual-recycled
Michelson interferometer. (b) Two coupled resonators. Inside dash line box, a single cavity is regarded as a mirror. (c)
The dual-recycled Michelson interferometer can be equivalent to two coupled resonators. \label{Fig1} }
\end{figure}

The schematic diagram is shown in Fig. 1(a). A continuous wave
laser feeds a dual-recycled Michelson interferometer, which is separated by a
50/50 beam splitter (BS) into two orthogonal directions and reflected back by two
mirrors ($M_{a1}$ and $M_{a2}$). Another two mirrors are placed on the
input path ($M_{2}$) and output path ($M_{0}$) of the Michelson
interferometer, respectively. This configuration is so called dual
recycling, i.e., the power (input path) recycling and the signal
(output path) recycling, which was introduced by Meers into the Michelson interferometer for reflecting the sideband back to the interferometer \cite{meers}. Thus the signal can be enhanced in the gravitational wave detection. The experimental demonstration was accomplished by Strain \cite{strain} in 1991. This configuration has been applied to the Advanced LIGO, which can significantly extend the detection range, improve the sensitivity and induce the first detection of detection of gravitational waves emitted from the merger of two black holes \cite{GW1,GW2,GW3}.

In this paper, it is employed to study the coupled-resonator induced transparency. For simplicity, the length from the 50/50 BS to the four end reflective mirrors have the same
optical length in our scheme. And the fine adjustment of the optical lengths or
the relative phases can be implemented by changing the voltage
of the piezoelectric transducers (PZT) mounted on the
mirrors. Consequently controlling the four mirrors can make the laser
resonant in the interferometer. At the same time, the relative
phases of two interferometer arms at
the 50/50 BS can be controlled. So the
reflectivity (or transmissivity) at the 50/50 BS can be adjusted.
Thus the system can be regarded as two coupled resonators whose coupling
strength is controllable, as shown in Fig. 1(b). For an usual configuration of two coupled resonators, a mirror with
adjustable transmissivity (or reflectivity) is placed in the middle
of the standing wave cavity. The coupling strength of the coupled
cavity mainly relies on the transmissivity of the middle mirror, usually which is difficult to be controlled or can not be changed. However, the
transmissivity (or the reflectivity) of the middle mirror can be
changed easily in our scheme because of the controllability of the interference of
two interferometer arms at the 50/50 BS. Therefore the transparent
window, the absorptive and the dispersive properties of the EIT-like
effect can be manipulated. Moreover, in principle it suits any laser
frequency in the coating band width of the mirrors. EIT effect is the Fano interference among the different transition pathways, which can be realized in quantum systems (such as in atom) or classic systems (such as coupled resonators). However, optical cavity, which is linear optical system, can be used as classic or quantum devices. EIT in the atomic system can be used to store the classical optical pulse or quantum field (such as squeezed light \cite{Honda,Appel}). Similarly, coupled-resonator-induced transparency can also be used to store the classical optical pulse or quantum field (such as squeezed light). These
characteristics make this system easily meet the requirements of the
quantum storage in the quantum information, such as for squeezed light \cite{Di}.

First, a single cavity is considered (for example, inside dash line box of Fig. 1(b)), whose reflected coefficient is written as
\begin{eqnarray}
\gamma_{1}&=&\frac{E_{out1}}{E_{in1}}=-\frac{r_{1}+r_{0}e^{2i\theta_{1}}}{1+r_{1}r_{0}e^{2i\theta_{1}}},
\end{eqnarray}
according to the input and output relationship. Here, $E_{in1}$ and $E_{out1}$ are input and reflected optical field. $r_{1}$ and $r_{0}$ are the
reflectivity of the input mirror $M_{1}$ and the other mirror $M_{0}$ respectively, where $r_{i}^{2}+t_{i}^{2}=1$ and $i\in\{0,1\} $. $\theta_{1}=2\pi\times(\omega+\Delta)T_{j}$  is the single-pass
phase shift of the intracavity field. Here $\omega$ and $\Delta$
are the resonant frequency and the frequency detuning of
the laser field respectively, $T_{j}=l_{j}/c$ is the single-pass time (the four arms have the same optical length, so we set $T_{j}=T$). Therefore a single cavity can be regarded as a mirror (as shown inside dash line box of Fig. 1(b)) with $M'_{1}(\gamma_{1},\mu_{1})$ and $|\gamma_{1}|^{2}+|\mu_{1}|^{2}=1$.
Then the reflected coefficient of two coupled resonators can be obtained by taking an iterative approach as shown in Fig. 1(b)
\begin{eqnarray}
\gamma_{2}&=&\frac{E_{out}}{E_{in}}=\frac{r_{2}+\gamma_{1}e^{2i\theta_{2}}}{1+r_{2}\gamma_{1}e^{2i\theta_{2}}}.
\end{eqnarray}

Now we can use the similar method to calculate the reflected coefficient of the dual-recycled Michelson interferometer. First, we consider a single cavity consists of the 50/50 BS with two interferometer arms and the signal recycling mirror $M_{0}$ as shown inside dash line box of Fig. 1(c). The two mirrors $M_{a1}$ and $M_{a2}$ of interferometer arms are high reflective. According to the input and output relationship
\begin{eqnarray}
E_{a1} &=& (-\emph{}E_{in1}+E'_{1})/\sqrt{2},\nonumber\\
E_{a2} &=&( E_{in1}+E'_{1})/\sqrt{2},
\end{eqnarray}
\begin{eqnarray}
E'_{a1} &=& -E_{a1}e^{2i\theta_{a1}} ,\nonumber\\
E'_{a2} &=& -E_{a2}e^{2i\theta_{a2}},\nonumber\\
E'_{1} &=& -r_{0}E_{1}e^{2i\theta_{1}}
\end{eqnarray}
and
\begin{eqnarray}
E_{out1} &=& (-E'_{a1}+E'_{a2})/\sqrt{2},\nonumber\\
E_{1} &=& (E'_{a1}+E'_{a2})/\sqrt{2},
\end{eqnarray}
we can obtain the reflected coefficient of the equivalent single cavity
\begin{eqnarray}
\gamma_{1}&=&e^{i(\theta_{a2}+\theta_{a1})}\frac{r_{1}+r_{0}e^{i(2\theta_{1}+\theta_{a2}+\theta_{a1})}}{1+r_{1}r_{0}e^{i(2\theta_{1}+\theta_{a2}+\theta_{a1})}},
\end{eqnarray}
where, $r_{1}=-\cos(\theta_{a1}-\theta_{a2})$. Thus, the 50/50 BS with two interferometer arms can be regarded as the middle mirror of two coupled resonators with the reflectivity of $r_{1}=-\cos(\theta_{a1}-\theta_{a2})$, which can be controlled easily by the relative phase between two interferometer arms. This relative phase can be manipulated very fast, for example through electro-optic modulator (EOM). Note that the extra phase $e^{i(\theta_{a2}+\theta_{a1})}$ is introduced in this configuration. Thus the free spectrum range of the equivalent single cavity is $(2l_{1}+l_{a1}+l_{a2})/c$ for the equivalent cavity 1 and $(2l_{2}+l_{a1}+l_{a2})/c$ for cavity 2. At last, the reflected coefficient of the dual-recycled Michelson interferometer is obtained by the iterative approach as shown in Fig. 1(c), which is the same as Eq. 2.

\begin{figure}
\centerline{
\includegraphics[width=3.3in]{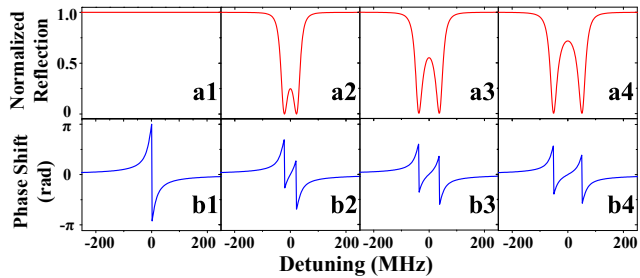}
} \vspace{0.1in}
\caption{(Color online) The theoretical results of the reflective spectrum and the corresponding phase shift at $M_{2}$ as the function of the laser frequency. (a1)-(a4), the reflective spectrum at $M_{2}$; (b1)-(b4), the phase shift of the reflective field. The single pass phases $\theta_{a1}=\theta_{a2}=2n\pi+\Delta T$ ($n=1,2...$) for (a1) and (b1), $\theta_{a2}=2n\pi-0.01\pi+\Delta T$ and $\theta_{a1}=2n\pi+0.01\pi+\Delta T$ for (a2) and (b2), $\theta_{a2}=2n\pi-0.015\pi+\Delta T$ and $\theta_{a1}=2n\pi+0.015\pi+\Delta T$ for (a3) and (b3), $\theta_{a2}=2n\pi-0.02\pi+\Delta T$ and $\theta_{a1}=2n\pi+0.02\pi+\Delta T$ for (a4) and (b4). Here, the single pass phases $\theta_{1}=\theta_{2}=2n\pi+\Delta T$, where $\Delta T$ corresponds to phase shift introduced by the laser detuning from the cavity resanance. Thus the figures in four columns correspond to the spectra with the reflectivity of the middle mirror $r_{1}=-1, -0.998, -0.996, -0.992$ in turn. Here, the wavelength $\lambda=1064$ nm, the optical lengths $l_{1,a1,a2,2}=28.196$ mm, the reflectivity of the four end mirrors $r^2_{0}=r^2_{2}=0.93$ and $r^2_{a1}=r^2_{a2}=1$. \label{Fig2} }
\end{figure}

EIT and Autler-Townes splitting have a similar transparency window in the absorption or transmission spectrum, despite the differences in their underlying physics. While the transparency window in EIT results from Fano interference among different transition pathways, in
Autler-Townes splitting it is the result of strong field-driven interactions leading to the
splitting of energy levels. All-optical analogues of
EIT and Autler-Townes splitting in coupled whispering-gallery-mode resonators \cite{lanyang} were studied by choosing the cavity parameters. Here, we don't focus on discrimination of two phenomena. EIT and Autler-Townes splitting can also be realized in the dual-recycled Michelson
interferometer by choosing the suitable cavity parameters. The theoretical calculation results of the EIT-like effect in the dual-recycled Michelson
interferometer are presented in Fig. 2. The coupling strength of the middle mirror can be expressed as \cite{ma}
\begin{eqnarray}
\Omega&=&\sqrt{1-r_{1}^2}\frac{c}{2l_{1}+l_{a1}+l_{a2}}.
\end{eqnarray}
The splitting of the spectra that depend on the coupling strength of the coupled cavity can be adjusted by controlling the interference of the optical field on the 50/50 BS, which are well explained above the theoretical analysis. Here, the reflective spectrum and the corresponding phase shift at $M_{2}$ are obtained from the magnitude and phase of the reflected coefficient $\gamma_{1}$ by tuning the frequency of injection laser with the wavelength $\lambda=1064$ nm. The optical lengths are $l_{1,a1,a2,2}=28.196$ mm and the reflectivity of the four end mirrors are $r^2_{0}=r^2_{2}=0.93$ and $r^2_{a1}=r^2_{a2}=1$. The phase difference $\theta_{a1}$ and $\theta_{a2}$ of two arms in Fig. 2 are 0, $0.02\pi$, $0.03\pi$ and $0.04\pi$, corresponding to the reflectivity of the middle mirror $r_{1}=-1$, $-0.998$, $-0.996$, $-0.992$, respectively (the coupling strength of the middle mirror $\Omega_{1}=0$, $2\pi\times26.6$ MHz, $2\pi\times39.8$ MHz, $2\pi\times53.1$ MHz according to Eq. 7). When we start to
set two cavities to be resonant simultaneously (that is increased coupling
strength), the single resonance split into two resonances whose
spectral distance (that is mode splitting) is increased (from the left to the left curve) with increasing
coupling strength.

\begin{figure}
\centerline{
\includegraphics[width=3in]{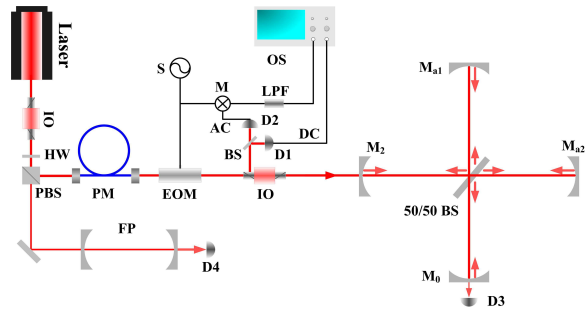}
} \vspace{0.1in}
\caption{(Color online). The schematic of experimental setup. IO, optical isolator; HW, half-wave plate; PBS, polarized beam splitter; PM, single-mode polarization-maintaining fiber; FP, Fabry-Poret cavity; EOM, electro-optic modulator; BS, beam splitter; S, signal generator; M, mixer; LPF, low-pass filter; D1-4, detectors; OS, oscilloscope. The power of the infrared laser before the interferometer is 10 mW.
\label{Fig4} }
\end{figure}

\begin{figure}
\centerline{
\includegraphics[width=3.5in]{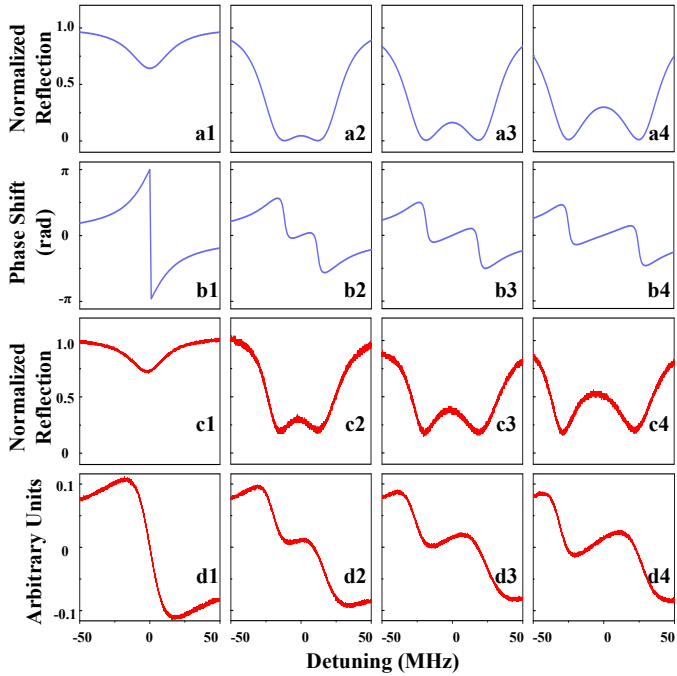}
} \vspace{0.1in}
\caption{(Color online). The reflection spectra and phase shifts of the dual-recycled Michelson
interferometer, as a function of the frequency detuning of the injected laser for different coupling strength of the equivalent middle mirror. (a1)-(a4) and (b1)-(b4) are the reflection spectra and the phase shift of the theoretical calculation results. Here, the intracavity loss is $0.4\%$ and the phase differences of two interferometer arms are 0, $0.0202\pi$, $0.0306\pi$, $0.0392\pi$. (c1-c4) and (d1-d4) are the experimental datum. The corresponding coupling strength are $\Omega=0$, $2\pi\times26.9$ MHz, $2\pi\times40.7$ MHz, $2\pi\times52.6$ MHz, respectively.
\label{Fig5} }
\end{figure}

The experimental setup is shown in Fig. 3. A diode-pumped continuous-wave ring single-frequency laser
provides the infrared light of 260 mW at 1064 nm. The laser is divided into two parts after an
optical isolator. One part is injected into a confocal Fabry-Perot (F-P) cavity (the free spectrum range
and the linewidth are 1.5 GHz and 10 MHz respectively, the finesse is 150) to monitor the laser frequency.
The other is coupled into a single-mode polarization-maintaining fiber to clean the spatial
modes of the laser. The output of the fiber passes through the electro-optic modulation with modulation frequency of
114.5 MHz and another optical isolator, then is injected into the dual-recycled Michelson
interferometer. A 50/50 BS is oriented at 45 degrees to the
laser beam. The power recycling mirror $M_{2}$ and the signal recycling mirror $M_{0}$ have the
reflectivity of 93$\%$ at 1064 nm, while $M_{a1}$ and $M_{a2}$ are the highly reflective mirrors for 1064
nm. All the four mirrors have radius of curvature of 30 mm and are all mounted on PZT to tune the
length of cavity subtly. The optical lengthes from the 50/50 BS to the four end reflective mirrors are about $28$ mm. In
fact, it is difficult to make the four arms completely same. But the optical length difference is
very small and it does not influence on the reflective spectra for several spectrum range. The free spectrum range of the single cavity is 2.66 GHz. The reflected field from the mirror $M_{2}$ of the dual-recycled Michelson interferometer is detected at the reflective window of the optical isolator 2. The reflected field from the optical isolator 2 is divided into two parts. One part is measured by detector 1 and its signal as the
reflection spectrum is directly shown on the oscilloscope (OS). And the other part is detected by detector 2 with a broad
detection bandwidth, and the output signal is combined with the
local signal at a mixer and the low-frequency mixed signal as the phase shift of the EIT-like
effect is collected by OS. This method corresponds to the Pound-Drever-Hall technique to obtain the error signal.

The measurement results are plotted in Fig. 4 (c1)-(c4) and (d1)-(d4) when the laser frequency is scanned. The voltages on the PZT's are controlled by four high voltage amplifier to make the laser resonant with the system and to
control two arms interference phases of the intracavity fields. In Fig. 4(c1), a dip appears in the reflective spectrum, which corresponds to the absorption profile of the equivalent single cavity since the two interferometer arms generate the constructive interference for the input mirror side at the 50/50 BS. A transparent window appears in the middle of the absorption profile as shown in Fig. 4(c2) when adding the coupling with the second cavity by controlling the relative phase of the two interferometer arms. And this transparent window becomes broader as shown in Fig. 4(c3) and (c4) when the coupling strength is increased further. Fig. 4(d1)-(d4) are the corresponding phase shift. The experimental results clearly display the coupled-resonator-induced transparent phenomena in the dual-recycled Michelson interferometer. Fig. 4(a1)-(a4) and (b1)-(b4) show the theoretical plots according to the experimental parameters, which are good agreement with the experimental results. Here, the intracavity loss is considered with $0.4\%$ in the theoretical calculation.

\begin{figure}
\centerline{
\includegraphics[width=3.5in]{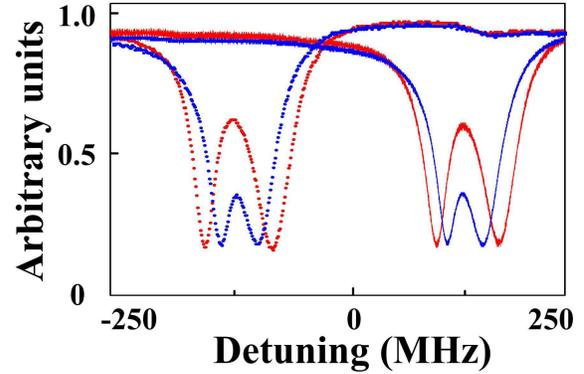}
} \vspace{0.1in}
\caption{(Color online). The reflective spectrum as we change the center frequency of the transparent window and the coupling strength. The coupling strength is $\Omega=2\pi\times81.2$ MHz for the dark line, $\Omega=2\pi\times41.8$ MHz for the gray line respectively. The center frequency of the transparent window is shifted $+125$ MHz for the solid line and $-125$ MHz for the dotted line.
\label{Fig7} }
\end{figure}

Furthermore, this system also displays more degrees of flexibility. The center frequency of the transparent window can be manipulated by the length from the 50/50 BS to the four end reflective mirrors. The center frequency of the transparent window is shifted $+125$ MHz for the solid line and $-125$ MHz for the dotted line, as shown in Fig. 5.

In conclusion, we study the coupled-resonator-induced transparency in the Michelson interferometer with dual-recycled configuration. This system is equivalent to the adjustable coupled resonator. The absorption and dispersion properties can be controlled by manipulating two interferometer arms. This work takes the first step of manipulating optical field
using the dual-recycled Michelson interferometer and provides a scheme for
the future studies on slow light, storage and retrieval. The basic requirements of a light-stopping process (capture,
storage, and release of the light pulse) are that the coupled resonator
system supports a large-bandwidth state to accommodate the input pulse bandwidth, which is then
dynamically tuned to an narrow-bandwidth state to stop
the pulse and done reversibly after some storage time to
release light pulse. The current structure of dual-recycled Michelson interferometer in our experiment can satisfy this requirement
completely to tune the cavity bandwidth dynamically.
We also hope that this work will stimulate the
research of manipulating the quantum optical fields, such as squeezed light \cite{Di}, which is similar to store the squeezed light in atomic vapor \cite{Honda,Appel}.

$^{\dagger}$ Corresponding author email: jzhang74@sxu.edu.cn, jzhang74@yahoo.com

\begin{acknowledgments}
This research was supported in part by the MOST (Grant No. 2016YFA0301602), National Natural Science Foundation
of China (NSFC) (Grant No. 11234008, 11361161002, 615712768, 11654002, 11804206), Natural Science Foundation of Shanxi
Province (Grant No. 2015011007), Research Project Supported by Shanxi Scholarship Council of China
(Grant No. 2015-002).
\end{acknowledgments}

\end{document}